# Organic light-emitting diodes using open-shell molecule as emitter: the emission from doublet

*Qiming Peng, Hongwei Ma, Youchun Chen, Chuanyou He, Ablikim Obolda, and Feng Li\**

State Key Laboratory of Supramolecular Structure and Materials, Jilin University
Qianjin Avenue, Changchun, 130012, P. R. China
\* E-mail: lifeng01@jlu.edu.cn

Organic light emitting diodes (OLEDs) have been expected to be the next flat displays and lighting sources due to their special features, such as easy processing, lightweight, and flexibility.[1,2] During the evolution of OLEDs over past two decades, many fluorescent and phosphorescent materials have been developed.[1,2,5-10] Although fluorescent materials have the advantage of cheapness, the upper limit of internal quantum efficiency (IQE) of OLEDs using them is only 25% according to the spin statistics.[3,4] In contrast, the IQE of OLEDs using phosphorescent materials has achieved almost 100%,[5-7] but practically useful phosphorescent materials are concentrated to the expensive Ir and Pt complexes. Recently, thermally activated delayed fluorescence (TADF) is proposed to harvest both singlet and triplet excitons in fluorescent OLEDs,[8] and very high efficiency in this kind OLEDs has been reported.[9] However, we note that the TADF-based OLED is in its young age and highly efficient TADF materials are few.

In this wok, different from the fluorescent, phosphorescent and TADF materials which are all closed-shell molecules, an open-shell organic molecule was used as the emitter of OLEDs. For a closed-shell molecule at ground state, there are two electrons in the highest occupied molecular orbital (HOMO). After the molecule is excited, one electron stays in the HOMO and another transits to the lowest unoccupied molecular orbital (LUMO). The spin configuration of the two electrons is either singlet or triplet, as shown in Scheme 1 (a).[11] According to the Pauli exclusion principle, the transition of triplet exciton to the ground state is forbidden.[3] However, if the emitting materials are open-shell molecules, the problem of triplets could be bypassed. The mechanism is as following: for an open-shell molecule, there is only one electron in the highest molecular orbital, i.e., the singly occupied molecular orbital (SOMO).[12] When this electron is excited into LUMO, the SOMO is empty, as shown in Scheme 1 (b). Thus, its transition back to the SOMO is allowed in despite of its spin state. So the upper limit of IQE of the OLEDs using open-shell molecules is theoretically 100%. Because one electron has two spin states, the excited state of open-shell molecules is called doublet.

Neutral radicals are one typical kind of open-shell molecules.[14] Generally, neutral radicals are quite unstable. However, through the molecular design, the stable neutral radicals can be obtained and they are able to withstand oxygen and light for an extremely long time at room temperature.[15] The neutral radicals are commonly studied in the fields of photophysics,[13,16-18] molecular magnetic materials,[19,20] spintronics,[21] and etc., due to the unpaired electron can easily take part in physical processes and chemical reactions.[14] Here, for the first time, we demonstrate a stable neutral π radical, α,γ-bisdiphenylene-β-phenylallyl (BDPA), as the emitter in OLEDs. The chemical structure of the BDPA is shown as the inset of Fig. 1 (a). For BDPA, an unpaired π electron delocalizes over the radical backbone in the SOMO (see the contour plot of SOMO in Fig. 1 (b)). The energy gaps between the HOMO, SOMO, and LUMO have been studied by Mullegger and co-workers.[15] We identified the energy level of SOMO of BDPA to be 5.6 eV via the ultraviolet photoemission spectroscopy (UPS) measurement, as can be seen in Fig. 1 (a). Then we obtained the energy levels of HOMO and LUMO according to the energy gaps from Mullegger, et. al., as shown in the left panels of Fig. 1 (b). The right panels of Fig. 1 (b) show the contour plots of the molecular orbits calculted by the density functional theory (DFT) using Gaussian 09 series of programs.[22]

The absorption (Abs.) and PL spectra of BDPA in tetrahydrofuran (THF) solution are shown in Fig. 2. From the Abs. spectrum, we can find a band peaking at 486 nm and another band peaking at 377 nm, which should correspond to the electronic transition from SOMO to LUMO and HOMO to LUMO, respectively. The shorter-wavelength portion of the Abs. spectrum is attributed to the transitions from SOMO/HOMO to higher energy levels. In view of this, we used two excitation sources (486 nm laser and 318 nm monochromatic light) to excite the sample, respectively. One band peaking at 541 nm can be seen under the excitation of 486 nm laser, which is ascribed to the electronic trasnsition from LUMO to SOMO. While when we excited the sample using a laser of 318 nm, two bands exist. The band peaking at 546 nm should be the transition from LUMO to SOMO, the other band peaking at 415 nm is attributed to the transition from LUMO to HOMO.

Further, we measured the lifetime of the excited states of BDPA (detected at 540 nm), as shown in Fig. 3. The weighted-average lifetime was calculated to be 3.26 ns, which is much shorter than that of the TADF emitters[9,23] ranging from microseconds to milliseconds. The long lifetimes may prevent the application of TADF emitters in displays due to the long response time of device.

We then fabricated the OLEDs using BDPA as the emitter. The structure of the OLEDs is indium tin oxide (ITO) / 1,3-bis(9-carbazolyl) benzene (mCP, 10nm) / emitting layer (50 nm) / tri-(8-hydroxyquinoline)-aluminum ($Alq_3$, 10 nm) / Lithium fluoride (LiF) (0.8 nm) /Aluminum (100 nm). The emitting layer is BDPA doped into $Alq_3$ with a doping concentration of wt. 10%. The layer of mCP functions as both transporting holes and blocking electrons. The electroluminescence (EL) spectrum is shown in Fig. 2. As can be seen, there is only one emission band with peak at 573 nm, indicating the emission comes from the trasnition from LUMO to SOMO. However, the EL Band is red-shifted and broader compared to the counterpart of PL spectrum, probably due to the formation of dimers and trimer of BDPA.[15]

Fig. 4 (a) shows the photographs of BDPA-based OLEDs at luminance of 10, 100, and 1000 $cd/m^2$, respectively. The current density-voltage-luminance (J-V-L) characteristics of the OLEDs are shown in Fig. 4 (b). As can be seen, the maximum luminance can reach 4879 $cd/m^2$. The maximum external quantum efficiency (EQE) of the OLED is 0.31% (see the inset of Fig. 4(b)), which is low. However, we note that the device performance of the BDPA-based OLED is not unsatisfactory as compared to that of the first reported Fluorescence-, Phosphorecence- and TADF-based OLEDs (Table 1).[1,5,6,8]

For an OLED, the ratio $\chi$ of light-emitting excitons (singlets in the Fluorescence- and TADF-based OLEDs) to the total excitons can be calculated from equation (1).[24]

$$\Phi = \chi \phi_{pl} \eta_r \eta_{out} \qquad (1)$$

Here $\Phi$ is the external quantum efficiency of the OLED, $\phi_{pl}$ is the PL efficiency of the emitter, $\eta_r$ is the fraction of injected charge carriers that form excitons, $\eta_{out}$ is the light out-coupling efficiency. According to the emission mechnism of open-shell molecules, $\chi$ shoule be 100 %. However, the $\chi$ was calculated to be 86 % by using the $\phi_{pl}$ of BDPA as 1.79 % in THF solution, and assuming a $\eta_{out}$ of 20 % and a $\eta_r$ of 1. The deviation might be resulted from the improperly using of 1 as the $\eta_r$, because the leakage current is unavoidable in an actual device.

In summary, we have fabricated OLEDs using a stable neutral π radical, BDPA, as the emitter. There is only one electron in the SOMO of this open-shell molecule. This feature makes the excited state of open-shell molecules be neither singlet nor triplet, but doublet. The key issue of how to harvest the triplet energy in an OLED is thus bypassed, due to the radiative decay of doublet is totally spin allowed. In the BDPA-based OLED, the emission was confirmed to be from the electronic transition from LUMO to SOMO, via the frontier molecular orbital analysis combined with the spectroscopy measurements. The quantum efficiency is currently not high, owing to the low PL efficiency of BDPA. Nevertheless, using neutral π radicals as emitter to fabricate OLED paves a new way to obtain 100% IQE of OLEDs, and finding neutral π radical with high emission efficiency is the next important issue.

*Experimental*

The neutral π radical BDPA complex with benzene (1:1) was purchased from Sigma Aldrich Corporation. Before evaporation, the material was degassed for several hours at 373 K under vacuum of $5\times10^{-4}$ Pa, insuring the benzene was removed. For UPS measurements, BDPA films (10 nm) were fabricated by thermal evaporation onto clean ITO coated glass substrates under high vacuum ($2\times10^{-4}$ Pa). Then the UPS spectrum was measured by a VG scienta XPS/UPS System under ultrahigh vacuum ($10^{-8}$ Pa). The resolution in the UPS experiment was less than 3 meV. The SOMO energy $E_{SOMO}$ was obtained by: $E_{SOMO}=E_{He-I} - E_{sec} + E_{onset}$. Here $E_{He-I}$ is the energy of photons from the He-I (21.2 eV) discharge lamp, $E_{sec}$ is the secondary electron cut off energy, and $E_{onset}$ is the SOMO onset energy. The DFT calculations were performed with the Gaussian 09 series of programs using the B3LYP hybrid functional and 6-31G(d) basis set. For the Abs. and PL measurements, BDPA was dissolved into the THF solution with a concentration of $5\times10^{-4}$ mol/L. Then the spectra were measured using a UV-Vis spectrophotometer (shimadzu UV-2550) and a spectrofluorophotometer (shimadzu RF-5301PC), respectively. The OLEDs were fabricated by the multiple source organic molecular beam deposition method at $2\times10^{-4}$ Pa. The current density–voltage ( J–V ) characteristics were measured by a Keithley 2400 source meter. The luminance–voltage ( L–V ) characteristic and the EL spectrum were measured by a PR650 spectroradiometer. For the exciton lifetime and PL efficiency measurements, an Edingburg fluorescence spectrometer (FLS980) was used.

*Acknowledgements*

We are grateful for financial support from the National Natural Science Foundation of China (grant numbers 61275036, and 21221063) and Graduate Innovation Fund of Jilin University (Project No. 2014012).

**Table 1**. Device performance of the first reported Fluorescence-, Phosphorecence- and TADF-based OLEDs and our neutral π radical-based OLED.

|  |  | $U_{Turn-on}$ (V) | $EQE_{Max}$ (%) | $L_{Max}$ (cd/m$^2$) | Ref. |
|---|---|---|---|---|---|
| Fluorescence- based OLEDs | | 2 | ~ 1 | > 1000 | [1] |
| Phosphorecence-based OLEDs | Os complex | 8 | < 0.1 | Very low | [5] |
|  | Pt complex | - | 4 | ~ 400 | [6] |
| TADF -based OLEDs | | 10 | - | - | [8] |
| Our neutral π radical-based OLED | | 4 | 0.31 | 4879 | |

$U_{Turn-on}$, the turn on voltage; $EQE_{Max}$, the maximum external quantum efficiency; $L_{Max}$, the maximum luminescence.

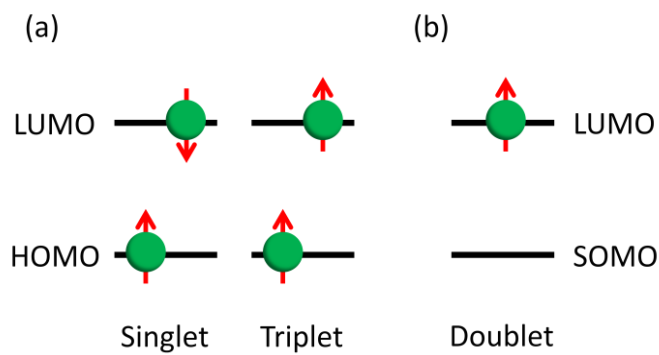

**Scheme 1.** Schematic diagram of the spin configuration of excited states. (a) For closed-shell molecules, the spin configuration of the excitons can be either singlet or triplet. (b) For open-shell molecules, the spin configuration is doublet.

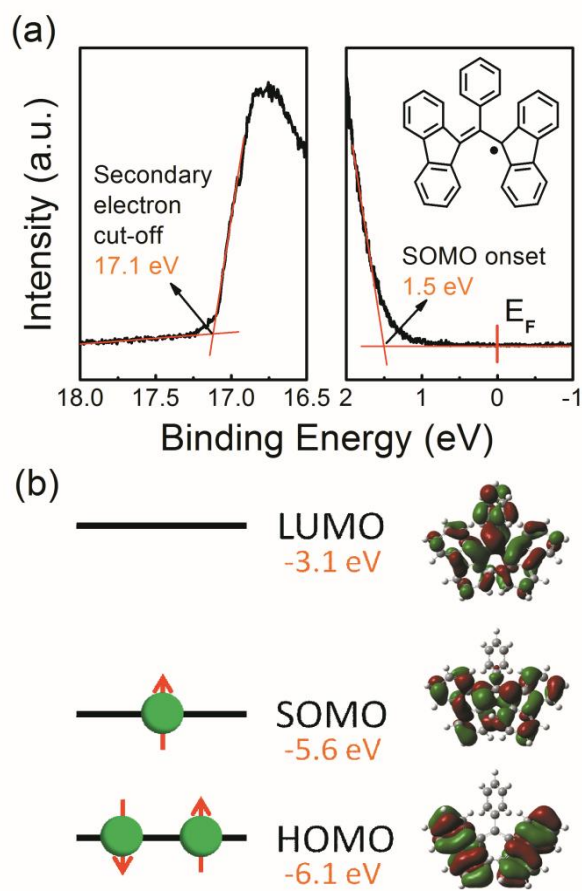

**Fig. 1.** (a) UPS spectrum of the BDPA, the inset shows the chemical structure of BDPA. (b) The energy levels (left panels) and contour plots (right panels) of the molecular orbits (LUMO, SOMO, and HOMO) of BDPA.

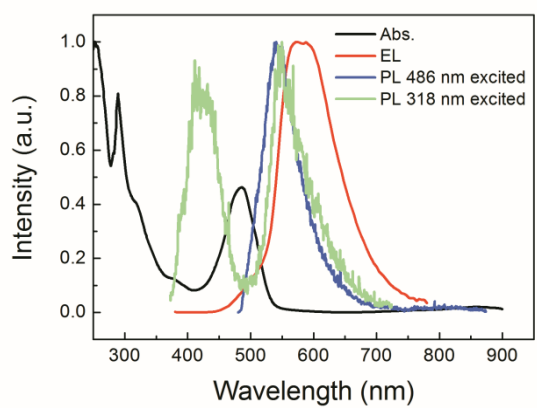

**Fig. 2.** The Abs. (black line) and PL (blue and green lines) spectra of BDPA in THF solution. The EL spectrum (red line) of the BDPA-based OLEDs.

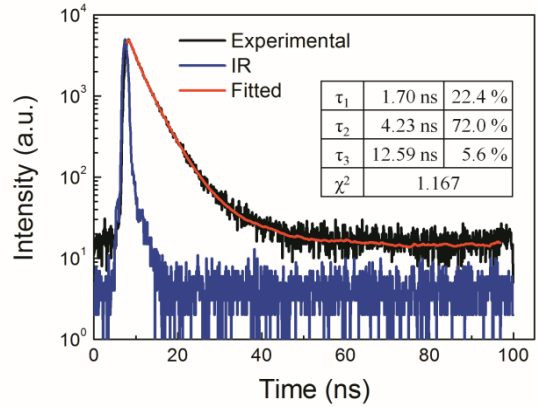

**Fig. 3.** The Experimental (black line) and fitted (red line) lifetimes of the BDPA in THF solution (detected at 540 nm). The blue line is the Instrument Response (IR). The weighted-average lifetime was calculated to be 3.26 ns by: $1/\tau = A_1/\tau_1 + A_2/\tau_2 + A_3/\tau_3$, where A is the percentage of each lifetime.

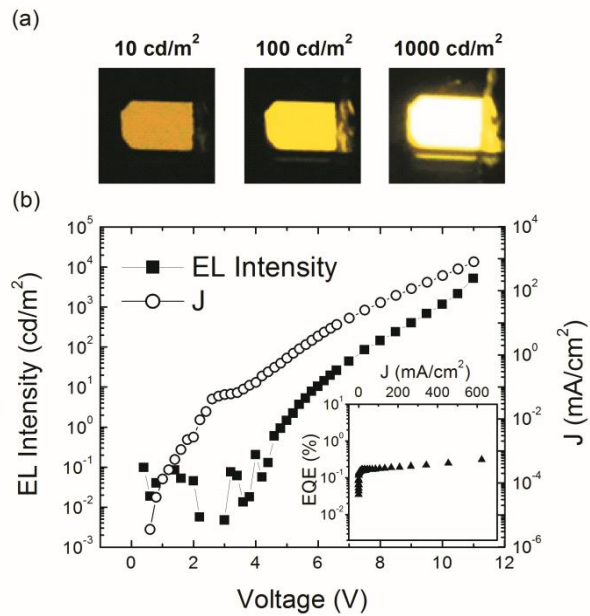

**Fig. 4.** (a) The photographs of BDPA-based OLEDs at luminance of 10, 100, and 1000 cd/m², respectively. (b) The J-V-L characteristics of the OLEDs. The inset shows the EQE as a function of current density.